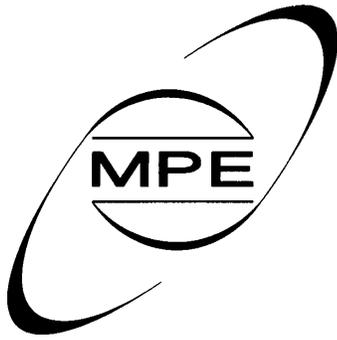

## Max-Planck-Institut
## Für Extraterrestrische Physik

# ROSAT observations of the radio and
# gamma-ray pulsar PSR 1706-44

W. Becker[1], K.T.S. Brazier[2] & J. Trümper[1]

[1] Max-Planck-Institut für extraterrestrische Physik
85740 Garching bei München, Germany

[2] Institute for Astronomy, University of Edinburgh,
Blackford Hill, Edinburg, EH9 2HJ

# ROSAT observations of the radio and gamma-ray pulsar PSR 1706-44

**W. Becker[1], K.T.S. Brazier[2], and J. Trümper[1]**

[1] Max-Planck-Institut für extraterrestrische Physik, 85740 Garching bei München, Germany

[2] Institute for Astronomy, University of Edinburgh, Blackford Hill, Edinburgh, EH9 3HJ, U.K.

**Abstract.** We report on the detection of PSR 1706-44 in two ROSAT-PSPC observations. The recorded source counts are unpulsed with a $2\sigma$ pulsed fraction upper limit of 18%. Spectral analysis did not distinguish between black-body and power law models; however, we argue that the lack of pulsations and the similarity in the pulsar's spin parameters to those of the Vela pulsar favour a power law model $dN/dE \propto E^{-2.4\pm0.6}$ and indicate synchrotron emission from a pulsar-powered nebula as the origin of the detected X-radiation. The X-ray flux derived for the power law model is $f_x = 3.2^{+6.5}_{-1.8} \times 10^{-12} \mathrm{erg/s/cm^2}$ within the 0.1–2.4 keV energy range. An upper limit for the neutron star's surface temperature is put at $\log T_s^\infty \sim 6.03$ K for a neutron star with a medium stiff equation of state (FP-model with $M = 1.4$ M$_\odot$ , $R = 10.85$ km). Slightly different values for $T_s^\infty$ are computed for the various neutron star models available in the literature, reflecting the difference in the equation of state. No soft X-ray emission is detected from the supernova remnant G343.1-2.3, proposed to be associated with PSR 1706-44.

## 1. Introduction

The launches in 1990 and 1991 of the X-ray satellite ROSAT and the *Compton* Gamma Ray Observatory CGRO began a new era for neutron star and pulsar research. At the beginning of these two experiments, only the Crab pulsar was known to emit pulsed radiation in both the X- and gamma ray wavebands; pulsed X-ray emission was known for just two other pulsars (PSR 1509-58 and PSR 0540-69) and only the Vela pulsar was seen to emit pulsed radiation in the gamma ray waveband like the Crab. Theoretical models to explain pulsar emission mechanisms and high-energy radiation characteristics were therefore based on the special findings for these few young pulsars and were hindered by the lack of empirical input. Now, with the superior data from the ROSAT and CGRO satellites, the number of rotation powered pulsars detected in both the X- and gamma ray wavebands has increased to six. PSR 1706-44 is one of them. Others are the Crab pulsar, PSR 1509-58 in the supernova remnant MSH 15-52 (detected only up to 2 MeV), the Vela pulsar, PSR 1055-52 and Geminga. Although this is still only about 1% of the known rotation powered pulsars, it is sufficient for a first study of the collective X- and γ-ray emission properties and a search for similarities in the relevant pulsar parameters (c.f. Becker 1994, Ögelman 1994, Ulmer 1994).

In this paper, we report on the ROSAT observations of PSR 1706-44. Soft X-ray emission from this pulsar has been detected recently in a serendipitous ROSAT observation by Becker et al. (1992), while the pulsed γ-radiation has been detected above 100 MeV using the Energetic Gamma-Ray Experiment Telescope (EGRET) aboard the Compton Gamma-Ray satellite CGRO (Thompson et al. 1992). The pulsar has a rotation period of 102 ms and a period derivative of $\dot{P} = 92.2 \times 10^{-15}$ s/s, implying a characteristic age of only $P/2\dot{P} \sim 17,400$ years and a rotational energy loss of $\dot{E} = 3.4 \times 10^{36}$ erg/s. With a dispersion measure based distance of $\sim 1.8$ kpc (Taylor & Cordes, 1993), the pulsar's spin-down energy flux density at Earth is $\dot{E}/4\pi d^2 = 8.6 \times 10^{-9}$ erg cm$^{-2}$s$^{-1}$, placing it fifth after the Crab, Vela, Geminga and the millisecond pulsar PSR J0437-4715 in a population of over 560 known rotation-powered pulsars. A possible association of PSR 1706-44 with the SNR G343.1-2.3 has been recently proposed on the basis of radio observations (McAdam et al. 1993). Distance inconsistencies as well as the lack of any recognizable interaction between the pulsar and the supernova shell, however, have led Frail et al. (1994) to question this association.

The structure of the paper is as follows: After describing the ROSAT observations in section 2, the results of a spatial analysis, including a search for X-ray point sources in the neighbourhood of PSR 1706-44 and for diffuse X-ray emission from the supernova remnant G343.1-2.3, are given in §3. The analysis of the photon arrival times and the spectral analysis are detailed in §4 and §5. The conclusions are presented in §6.

## 2. ROSAT Observations

In the course of the ROSAT mission, PSR 1706-44 was observed with the ROSAT PSPC (Position Sensitive Proportional Counter) on three separate occasions. The first of these was during the all-sky survey, in which the pulsar position was observed for only $\sim 150$ seconds. PSR 1706-44 was not detected on that occasion. The following analysis is therefore based on two subsequent pointed observations of $\sim 3400$ and $\sim 8400$ seconds which took place on 24th-25th February and 22nd-25th September 1992. The target of the February observation was the low-mass X-ray binary system 4U 1705-44 and PSR 1706-44 was seen serendipitously at an off-axis angle of 23 arcminutes, where the telescope's angular resolution is somewhat lower than on-axis. The September observation, centred at the pulsar position, provides therefore the most useful information. Details of the two PSPC pointings including the start, stop and effective exposure times are given in Tab. 1. Also listed are the number of source counts determined by the maximum likelihood analysis described in section 3.1.

**Table 1.** ROSAT pointed observations of PSR 1706-44

| ROSAT | start time (UTC) | stop time (UTC) | exposure time | source counts |
|---|---|---|---|---|
| PSPC AO-2 | 24. Feb. 1992 08:24:16 | 25. Feb. 1992 09:52:43 | 3381 s | $\sim 72$ |
| PSPC AO-3 | 22. Sep. 1992 05:32:36 | 25. Sep. 1992 07:25:18 | 8411 s | $\sim 186$ |

## 3. Spatial Analysis

### 3.1. The X-ray counterpart of PSR 1706-44

In order to establish the ROSAT detection of PSR 1706-44 (c.f. Becker et al. 1992) and to determine the pulsar's X-ray position and count rate, a spatial fit to the data was obtained from a spline fit to the background level in combination with a maximum likelihood analysis of the source counts. The dead-time, vignetting- and background-corrected net count rate deduced from this analysis is $0.022 \pm 0.002$ cts/s for the energy range 0.1–2.4 keV. The source position was determined as $\alpha(2000) = 17^h 09^m 42\overset{s}{.}33\ (\pm 0\overset{s}{.}5)$ and $\delta(2000) =$ $-44° 29' 11\overset{''}{.}4\ (\pm 8^s)$, where the errors are dominated by uncertainty in the satellite pointing direction. The source location is consistent with the pulsar's VLA position recently reported by Frail, Goss & Whiteoak (1994) and the timing position computed by Johnston et al. (1994). No other likely counterpart was found within the ROSAT error box. The source extent is in agreement with the PSPC point spread function.

### 3.2. Other X-ray sources in the field of view

Besides the bright LMXB 4U 1705-44 (Predehl & Schmitt, 1994), two other X-ray sources are detected in the vicinity of PSR 1706-44. Using the ROSAT source name convention, RX J1710.7-4433 is located at $\alpha(2000) = 17^h 10^m 42\overset{s}{.}5\ (\pm 0\overset{s}{.}5)$ and $\delta(2000) = -44^d 33^m 33\overset{s}{.}3\ (\pm 8^s)$. The net count rate of this source is $0.038 \pm 0.002$ cts/s within 0.1–2.4 keV. Inspecting the digitized UK Schmidt plates and the SIMBAD catalogue for an optical counterpart, we have identified the X-ray source as a fifth magnitude foreground star, HD 154948. A second, weaker X-ray source RX J1709.4-4429 was found $\sim 3.6$ arcmin from the pulsar position at $\alpha(2000) = 17^h 09^m 27\overset{s}{.}9\ (\pm 0\overset{s}{.}5)$, $\delta(2000) = -44^d 29^m 24\overset{s}{.}6\ (\pm 8^s)$, with a net count rate of $0.004 \pm 0.0008$ cts/s. A search for an optical counterpart resulted in three candidates, all of which are extragalactic. Neither of these sources interferes with the analysis of the pulsar. X-ray emission is not detected from the radio point source close to the geometric centre of the supernova remnant G343.1-2.3 (McAdam et al. 1993).

### 3.3. Search for diffuse X-ray emission from G343.1-2.3

Based on the $\Sigma$-D relation, McAdam et al. (1993) and Frail et al. (1994) estimated the distance of the $\sim 44 \times 32$ arcmin supernova shell G343.1-2.3 to be about 3–4 kpc. The origin of its radio emission is thought to be nonthermal. The pulsar lies about 23 arcminutes from the apparent geometric center of the SNR (c.f. Fig. 1), which for a pulsar of age 17,400 years implies a proper motion of $\mu = 0.08$ arc seconds per year (Frail et al. 1994). Experimental determination of the proper motion is the subject of an on-going programme (Johnston, private communication) and will eventually determine whether the pulsar and SNR are physically associated.

Figure 1 shows the pulsar/SNR field as a superposition of the Molonglo 843 MHz radio flux map from McAdam et al. (1993) and the X-ray intensity map from the September observation. There is no visible sign of a correlation between the diffuse X-ray emission and the supernova radio shell. A possible explanation is that X-ray emission from the SNR is inherently weak and is rendered undetectable by interstellar absorption. As we will show in §5, the column density towards the supernova remnant is relatively high: $N_H = 2$–$5 \times 10^{21}$ cm$^{-2}$. A further hindrance to the detection of extended emission from the remnant is the presence of the strong X-ray source 4U 1705-44, which dominates a wide area of the PSPC field and may bury diffuse emission from the remnant.

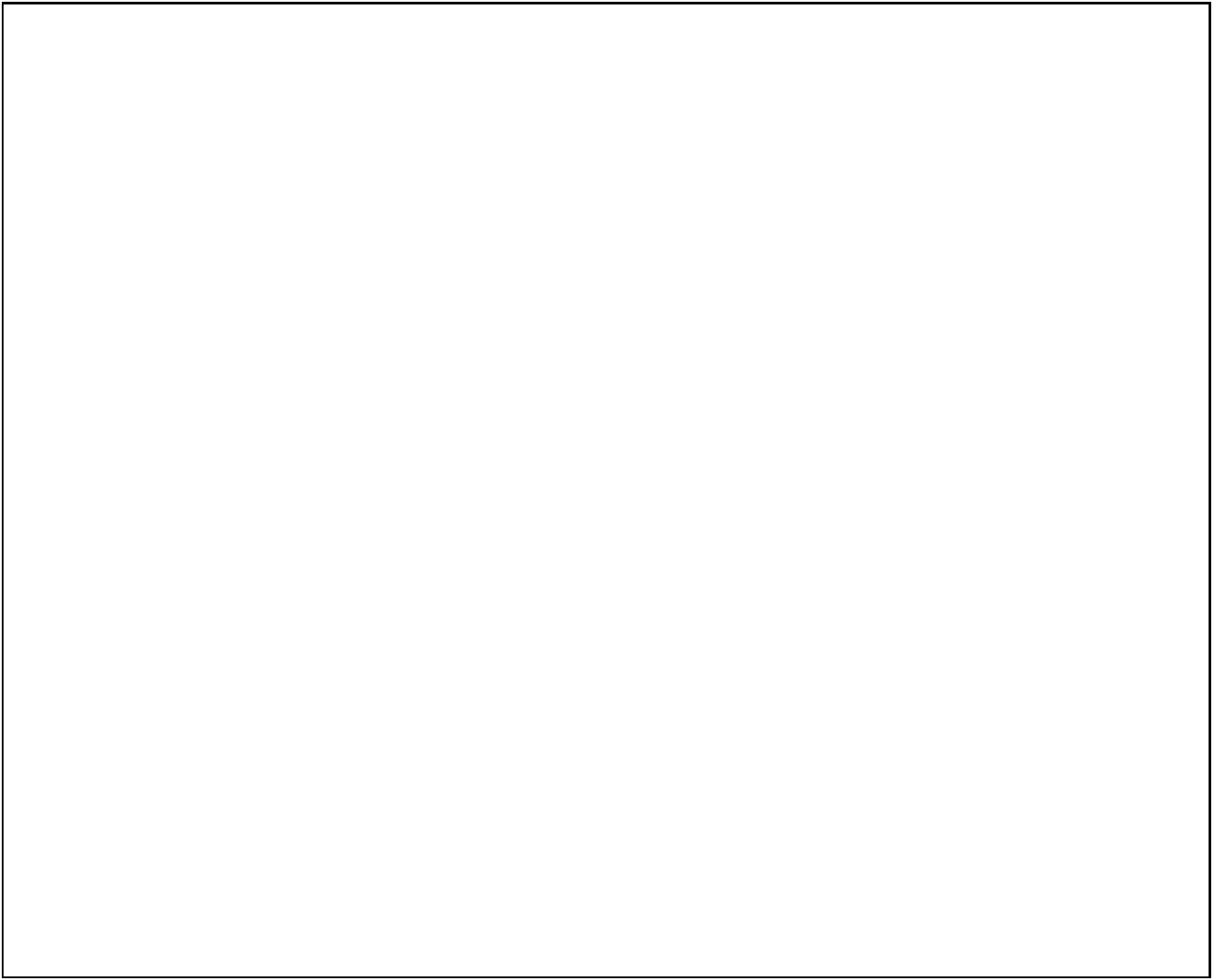

**Fig. 1.** Superposition of the 0.1–2.4 keV X-ray intensity map with the Molonglo 843 MHz flux density map of the SNR G343-2.3 region (contour lines; radio point sources have been removed). The position of the pulsar at the edge of the elliptical half-shell is marked by arrows. Contour lines are at (2,4,5.5,7,9,28,32,200) mJy per beam of 43" × 63". The bright X-ray source at the north is the low mass X-ray binary system 4U 1705-44. The contour lines cover the small X-ray source RX J1709.4-4429 close to the pulsar position. RX J1710.7-4433 is visible in south-east direction from the pulsar position.

To quantify the X-ray emission component from G343.1-2.3, we have computed a flux upper limit for that part of the supernova shell which is visible at 843 MHz with the Molonglo synthesis telescope (c.f. Fig. 1). The $2\sigma$ count rate upper limit determined for the radio shell is 0.018 cts/s. Assuming a power-law spectrum $dN/dE \propto E^{-\alpha}$ with photon-index $\alpha = 2$ and column density $N_H = 5 \times 10^{21}$ cm$^{-2}$, the count rate corresponds to an unabsorbed flux of $f_x \leq 9 \times 10^{-13}$ erg/s/cm$^2$ within 0.1–2.4 keV.

## 4. Timing analysis

To test the X-ray flux for a modulation at the pulsar's radio frequency, a photon arrival time analysis was applied to the photons from within a radius of 60 arcsec from the source position for the February data and 50 arcsec for the September data, respectively. These radii were chosen with respect to the point spread functions and include more than 99.9% of the source photons. About 10% of the selected photons are background photons. Because of the large gap between the two ROSAT observations and the occurrence of a glitch between 3rd May and 9th July 1992 (Johnston, private communication), the February and September data were analysed separately. Despite the glitch, valid pulsar ephemerides were available for each of the ROSAT observation epochs from the CGRO/radio timing database held at Princeton University (Johnston et al. 1994). The ephemeris used to analyse the September data is shown in Tab. 2.

**Table 2.** Radio ephemeris for PSR 1706-44, valid within the time period 48812–48928 MJD (Johnston et al. 1994)

| | |
|---|---|
| Right Ascension (J2000) | $17^h\ 09^m\ 42^s.16$ |
| Declination (J2000) | $-44°\ 28'\ 57"$ |
| Rotation frequency, $f$ | 9.7608828344790 Hz |
| Rate of frequency change, $\dot{f}$ | $-8.89786 \times 10^{-12}$ Hz/s |
| Reference epoch $t_{ref}$ | 2448861.5 (JD, TDB) |
| Phase of radio pulses at $t_{ref}$ | 0.9216 |

The pulsar position is the timing position valid for $t_{ref}$.

Using the standard procedure to barycentre and correct the ROSAT recorded photon arrival times (e.g. Becker et al. 1993), neither data set was found to contain a significant modulation at the pulsar's rotation period. A $2\sigma$ upper limit of 18% for the fraction of pulsed photons was computed by fitting a sine wave to the pulsar lightcurve deduced from the period-folded photon arrival times of the September data. The X-ray lightcurves of PSR 1706-44 from the February and September data, together with its $\gamma$-ray pulse profile, are shown in Fig. 2.

## 5. Spectral analysis

### 5.1. The soft X-ray spectrum of PSR 1706-44

Soft X-ray emission from PSR 1706-44 is expected on the basis of thermal emission from the neutron star's surface and from synchrotron or curvature radiation released by accelerated $e^{\pm}$-pairs moving along the curved magnetic field lines in the neutron star's magnetosphere (c.f. Michel 1991 and references therein). Further X-radiation may come from a pulsar powered synchrotron nebula and possibly even from the interaction between a relativistic pulsar-driven wind and the interstellar medium.

In order to investigate the spectral characteristics of the pulsar's X-ray emission, a study of PSR 1706-44 was conducted on the basis of the combined February and September data. The 256 energy channels of the PSPC detector were binned into eight energy bands so that the signal-to-noise ratio in each band was between four and five sigma. Model spectra were then compared with the observed spectrum. In total, three spectra were tested against the data: black-body, thermal bremsstrahlung and power law. There were too few source photons to determine which of the models was closest to the source spectrum, but all models produced acceptable fits to the data. Table 3 summarizes the best fit parameters for the three spectral models.

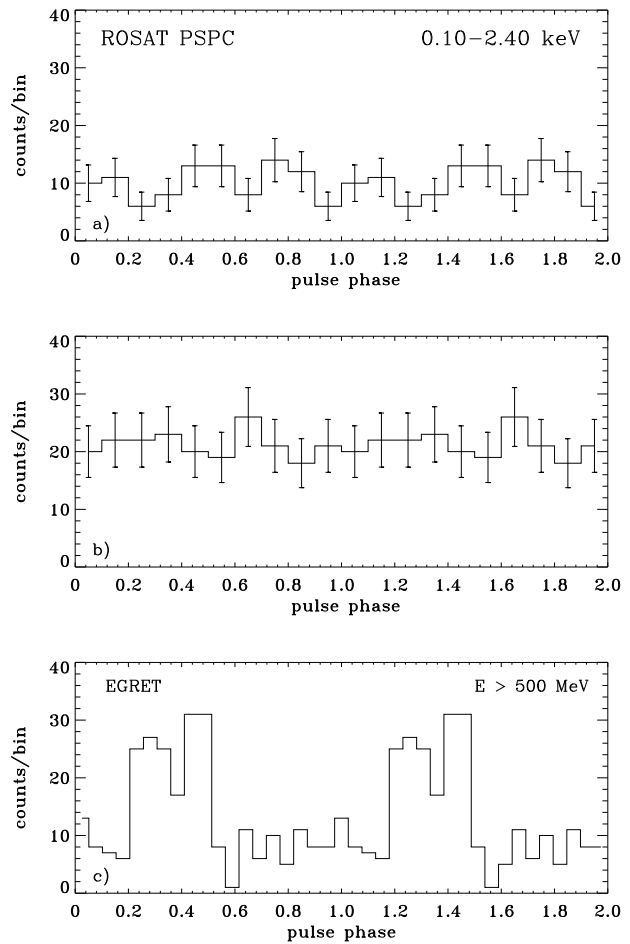

**Fig. 2. a and b** X-ray lightcurves of the 102 ms pulsar PSR 1706-44, based on the February (Fig. 2a) and September data (Fig. 2b). No significant modulations with the pulsar's radio period have been detected in the ROSAT data. **c)** The $\gamma$-ray lightcurve of PSR 1706-44 is characterized by a single pulse, contrary to the $\gamma$-ray pulse profiles of the Crab, Vela and Geminga pulsars which are double peaked with a phase separation of $\sim 0.4$ between the two peaks. Phase 0 in Fig. 2a corresponds to JD= 2448658.5 TDB while in Fig. 2b,c it corresponds to JD= 2448861.5 TDB. ($\gamma$-ray lightcurve adopted from Thompson et al. 1992).

Describing the data in terms of black-body emission implies a temperature of $T = 4.6 \pm 1.3 \times 10^6$ K, about three times that predicted by standard cooling scenarios for the surface of a $\sim 17,400$ year old isolated neutron star (Van Riper 1991, Umeda et al. 1993, Page 1994). Using $R = d/T^2 \sqrt{f_{bol}/\sigma}$ with $\sigma = 5.67 \times 10^{-5}$ erg cm$^{-2}$ s$^{-1}$ K$^{-4}$ and $f_{bol} = 7 \pm 0.3 \times 10^{-13}$ erg cm$^{-2}$ s$^{-1}$, we find $R = 0.15 \pm 0.1 \times d_{kpc}$ km for the radius of the emitting area, suggesting that the X-ray emission could be from a small, heated polar cap rather than being cooling radiation emitted from the whole neutron star surface. However, although the size of the emitting polar

**Table 3.** Spectral models of PSR 1706-44.

| Model | $N_H \times 10^{21}$ cm$^{-2}$ | Model parameter | Luminosity erg/s |
|---|---|---|---|
| Black-body | $2.5 \pm 1.2$ | $T = (4.6 \pm 1.3) \times 10^6$ °K | $\sim 2.7 \times 10^{32}$ |
| Therm. Bremsstr. | $4.6 \pm 1.3$ | $T = (1.9 \pm 1.4) \times 10^7$ °K | $\sim 5.2 \times 10^{32}$ |
| Power law | $5.4 \pm 1.5$ | Photon-Index = $2.4 \pm 0.6$ | $\sim 1.3 \times 10^{33}$ |

The pulsar luminosity has been calculated with a nominal pulsar distance of 1.8 kpc. In the case of the black-body spectrum, the luminosity is the bolometric luminosity. Radiation is isotropic in all cases.

cap zone and the temperature are in agreement with the results found for Geminga (Halpern & Ruderman 1993), PSR 0656+14 (Ögelman 1994), PSR 1055-52 (Ögelman & Finley 1993), PSR 1929+10 (Yancopoulos et al. 1994) and the millisecond pulsar PSR J0437-4715 (Becker & Trümper 1993), the model is in conflict with the timing results since polar cap emission is expected to be highly spin modulated, and viewing-angle effects cannot eliminate this conflict. We therefore favour an explanation in which the origin of the pulsar's soft X-ray emission is synchrotron and/or curvature radiation, released from regions outside the light cylinder. No modulation at the neutron star rotation frequency is then expected, and the spectrum should follow a power law: $dN/dE \propto E^{-\alpha}$. Considering the pulsed fraction upper limit of 18%, a small as yet undetected modulation from magnetospheric or surface emission may exist alongside the unpulsed flux. The spectral fit of the power-law model is shown in Fig. 3.

### 5.2. Neutron star surface temperature upper limit

If we accept that the majority of the observed photons are synchrotron radiation, the thermal contribution from the neutron star surface must be minor. An upper limit to the surface temperature $T_s$ can therefore be estimated as the temperature that would correspond to the whole of the observed X-ray flux radiated by a neutron star of radius $R$, at a distance $d$ and through an absorption column density $N_H$ (e.g. Becker et al. 1993). Due to the gravitational redshift $z$, only $T_s^\infty = T_s/(1+z)$ can be observed. Theoretical calculations of cooling rates, i.e., the rate of thermal emission from the surface of a neutron star of given age, depend mainly on the internal structure of the star, in particular on the equation of state at supernuclear densities. In the literature various neutron star models of different equations of state have been proposed. To make our results comparable with the predictions of these models, the stellar parameters $M$ (mass) and $R$ (radius) from these models have been used for computing $T_s^\infty$ and $L_\gamma^\infty$, respectively. Table 4 shows the results for two combinations of pulsar distance $d$ and column density $N_H$. Using the kinematic distance limits given in Koribalski et al. 1994, combined with the full range

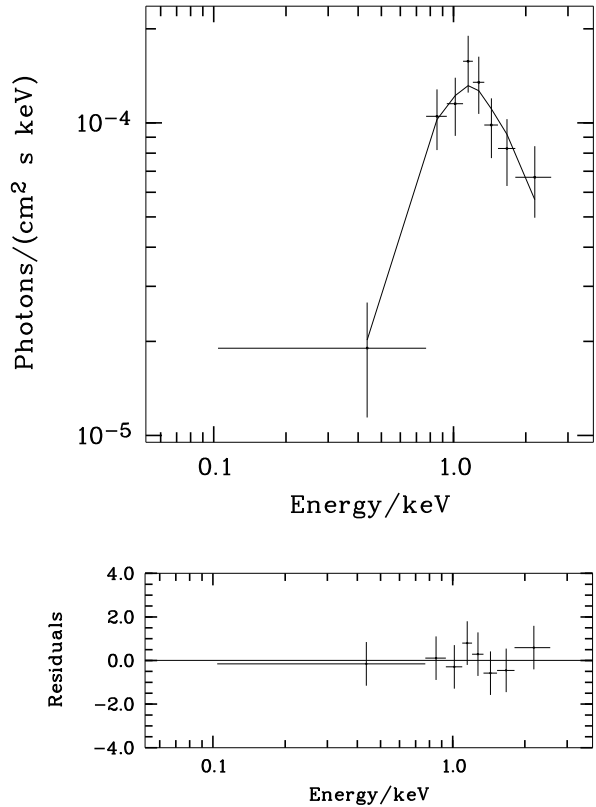

**Fig. 3.** Spectral fit of the pulsar's soft X-ray spectrum with a power-law $dN/dE \propto E^{-2.4}$. The source is barely visible below 0.5 keV, indicating a hard and absorbed spectrum. The residuals are in units of $1\sigma$.

of column densities inferred from the spectral analysis (i.e. $d = 2.7 \pm 0.9$ kpc and $N_H = 4.6^{+2.3}_{-3.3} \times 10^{21}$ cm$^{-2}$) implies temperature upper limits which are consistent with the predictions of standard cooling models. None of the various neutron star models is excluded by these temperature upper limits even if heating processes like frictional heating or crust cracking are taken into account. This situation changes, however, if we compute the surface temperature upper limits for the dispersion measure based distance of $d = 1.8 \pm 0.5$ kpc (Taylor & Cordes 1993) and the column density $N_H = 2.5 \pm 1.2 \times 10^{21}$ cm$^{-2}$

**Table 4.** Surface temperature Upper Limits for PSR 1706-44

| Model | Radius km | Mass $M_\odot$ | $d = 2.7 \pm 0.9$ kpc $N_H = 4.6^{+2.3}_{-3.3} \times 10^{21}$ cm$^{-2}$ | | $d = 1.8 \pm 0.5$ kpc $N_H = 2.5^{+1.2}_{-1.2} \times 10^{21}$ cm$^{-2}$ | |
|---|---|---|---|---|---|---|
| | | | $\log T_s^\infty$ K | $\log L_\gamma^\infty$ erg/s | $\log T_s^\infty$ K | $\log L_\gamma^\infty$ erg/s |
| PS[1] | 16.10 | 1.31 | $6.09^{+0.06}_{-0.14}$ | $33.76^{+0.23}_{-0.54}$ | $6.00^{+0.06}_{-0.08}$ | $33.40^{+0.23}_{-0.32}$ |
| PS[2] | 15.83 | 1.40 | $6.10^{+0.06}_{-0.14}$ | $33.76^{+0.23}_{-0.54}$ | $6.00^{+0.06}_{-0.08}$ | $33.39^{+0.23}_{-0.32}$ |
| MPA[3,4] | 12.45 | 1.40 | $6.12^{+0.06}_{-0.14}$ | $33.68^{+0.23}_{-0.54}$ | $6.02^{+0.06}_{-0.08}$ | $33.31^{+0.23}_{-0.32}$ |
| PAL33[4] | 11.91 | 1.40 | $6.12^{+0.06}_{-0.13}$ | $33.67^{+0.23}_{-0.54}$ | $6.03^{+0.06}_{-0.08}$ | $33.30^{+0.23}_{-0.31}$ |
| UV14[4,5] | 11.20 | 1.40 | $6.12^{+0.06}_{-0.13}$ | $33.65^{+0.23}_{-0.54}$ | $6.03^{+0.06}_{-0.08}$ | $33.28^{+0.23}_{-0.31}$ |
| UU[6] | 11.14 | 1.40 | $6.12^{+0.06}_{-0.13}$ | $33.65^{+0.23}_{-0.54}$ | $6.03^{+0.06}_{-0.08}$ | $33.28^{+0.23}_{-0.31}$ |
| PAL32[4] | 11.02 | 1.40 | $6.13^{+0.06}_{-0.13}$ | $33.64^{+0.23}_{-0.54}$ | $6.03^{+0.06}_{-0.08}$ | $33.27^{+0.23}_{-0.31}$ |
| FP[1] | 10.90 | 1.29 | $6.13^{+0.06}_{-0.13}$ | $33.63^{+0.23}_{-0.54}$ | $6.04^{+0.06}_{-0.08}$ | $33.26^{+0.23}_{-0.31}$ |
| FP[4,5] | 10.85 | 1.40 | $6.13^{+0.06}_{-0.13}$ | $33.64^{+0.23}_{-0.54}$ | $6.03^{+0.06}_{-0.08}$ | $33.27^{+0.23}_{-0.31}$ |
| AV14[4,5] | 10.60 | 1.40 | $6.13^{+0.06}_{-0.13}$ | $33.63^{+0.23}_{-0.54}$ | $6.04^{+0.06}_{-0.08}$ | $33.26^{+0.23}_{-0.31}$ |
| AU[6] | 10.40 | 1.40 | $6.13^{+0.06}_{-0.13}$ | $33.63^{+0.23}_{-0.54}$ | $6.04^{+0.06}_{-0.08}$ | $33.26^{+0.23}_{-0.31}$ |
| FP(pion)[1] | 9.40 | 1.27 | $6.14^{+0.06}_{-0.13}$ | $33.58^{+0.23}_{-0.53}$ | $6.05^{+0.06}_{-0.08}$ | $33.21^{+0.23}_{-0.31}$ |
| PAL(kaon)[7] | 8.20 | 1.40 | $6.15^{+0.06}_{-0.13}$ | $33.57^{+0.23}_{-0.53}$ | $6.05^{+0.06}_{-0.08}$ | $33.20^{+0.23}_{-0.31}$ |
| BPS[1] | 7.90 | 1.35 | $6.15^{+0.06}_{-0.13}$ | $33.55^{+0.23}_{-0.53}$ | $6.06^{+0.06}_{-0.08}$ | $33.18^{+0.23}_{-0.31}$ |
| BPS[2] | 7.35 | 1.40 | $6.15^{+0.06}_{-0.13}$ | $33.55^{+0.23}_{-0.53}$ | $6.06^{+0.06}_{-0.08}$ | $33.18^{+0.23}_{-0.31}$ |
| PAL(kaon)[7] | 7.00 | 1.40 | $6.15^{+0.06}_{-0.13}$ | $33.54^{+0.23}_{-0.53}$ | $6.06^{+0.06}_{-0.08}$ | $33.17^{+0.23}_{-0.31}$ |

The stellar parameters radius and gravitational mass are taken from: [1]Umeda et al. (1993), [2]Van Riper (1991), [3]Müther et al. (1987), [4]Page (1994), [5]Wiringa et al. (1988), [6]Chong & Cheng (1993), [7]Thorsson et al. (1994). The allowed range quoted for $T$ and $L$ includes the uncertainties of distance and column density (c.f. also Fig.4).

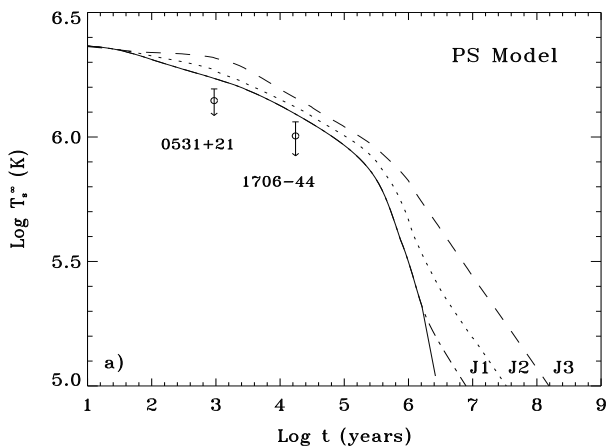

**Fig. 4.** Temperature upper limit for PSR 1706-44, using the PS model parameters from Umeda et al. (1993). Also shown are the cooling curves of this model and the temperature upper limit for the Crab pulsar (Becker & Aschenbach 1994). The upper limits are represented by a circle while the arrows display the uncertainty range due to the lack of exact values for distance and column density. The dotted and dashed lines show the predicted thermal evolution of the neutron star with internal frictional heating in the case of strong (J3), weak (J2) and superweak (J1) pinning of crustal superfluid vortex lines.

which is inferred both from the black-body fit (c.f. Tab. 3) and the dispersion measure $DM = 75.68$ pc/cm$^3$ (Taylor, Manchester & Lyne, 1993) by assuming a mean electron density of $\bar{n}_e \cong 0.03$ cm$^{-3}$. The surface temperature upper limit for PSR 1706-44 is then below the temperature predicted by the PS model, invalidating its application (c.f. Fig.4). The comparision with the FP model, however, still shows agreement with the upper limit but only if there is no strong frictional heating. A more detailed comparison between the different thermal evolution models and the temperature upper limits for the ROSAT detected rotation powered pulsars can be found in Becker (1994).

## 6. Conclusion

PSR 1706-44 is very similar in its spin parameters to the Vela pulsar ($P = 89$ ms, $\dot{P} = 125 \times 10^{-15}$ s/s), so it is interesting to compare their observed X-ray characteristics. The ~ 2 arcminute synchrotron nebula around the Vela pulsar has been known since the Einstein mission (Harnden et al. 1985). Pulsed soft X-ray emission from the pulsar was discovered more recently with ROSAT (Ögelman et al. 1993). The fraction of pulsed emission was estimated as ~ 4.5% of the combined emission detected from the pulsar and synchrotron nebula. The soft X-ray spectrum of the pulsed Vela emission is probably thermal and only visible below $\sim 1.2$ keV. The emission from the synchrotron nebula dominates the Vela pulsar spectrum above $\sim 0.5$ keV.

When the results for the Vela pulsar are compared with those for PSR 1706-44, they can be seen to be consistent with our proposal that the soft X-ray emission from PSR 1706-44 is dominated by the synchrotron radiation of an unresolved nebula. The thermal emission component, which for the Vela pulsar is only visible as pulsed X-ray emission, is rendered undetectable by the low photon statistics and the dominant role of the harder nebula emission. The photon-index of $\alpha = 2.4 \pm 0.6$ found for PSR 1706-44 is in agreement with the results found for the synchrotron nebula of PSR 1509-58 in MSH 15-52 (Becker 1994), Vela (Ögelman et al. 1993), the Crab pulsar (Becker 1994), PSR 1951+32 in CTB 80 (Safi-Harb & Ögelman 1994) and PSR 0540-69 (Finley et al. 1993).

In summary, the ROSAT data support the interpretation of PSR 1706-44 as a Vela-like pulsar. With the exception of the different gamma-ray pulse shapes, this is also indicated by its gamma-ray emission characteristics (Thompson et al. 1992). It will be interesting to see if the ASCA observations of PSR 1706-44 – which are scheduled for September 1994 – can put further constraints on this interpretation.

*Acknowledgements.* We thank Dr. Bruce McAdam and Dr. Taisheng Ye from Department of Astrophysics of the University of Sydney for providing us with the radio point source removed MOST image of the supernova remnant G343.1-2.3. The data were analysed using the software package EXSAS.

# References

Becker, W., Predehl, P., Trümper, J. & Ögelman, H.B., 1992, IAU Circ. 5554

Becker, W., Trümper, J. & Ögelman, H., 1993, in *Isolated Pulsars*, eds K.A. Van Riper, R.I. Epstein & C. Ho, 104, Cambridge University Press

Becker, W., Brazier, K.T.B. & Trümper, J., 1993, A&A, 273, 421

Becker, W., Trümper, J., 1993, Nat, 365, 528

Becker, W., Aschenbach, B., 1994, in *The Lives of Neutron Stars*, eds A. Alpar, U. Kiliźoğlu & J. van Paradijs, Kluwer Academic Publishers, in print

Becker, W., 1994, thesis, Ludwig-Maximilians-Universität München

Chong, N. & Cheng, K.S., 1993, ApJ, 417, 279

Finley, J.P., Ögelman, H., Hasinger, G. & Trümper, J., 1993, ApJ, 410, 323

Frail, D.A., Goss, W.M. & Whiteoak, J.B.Z., 1994, submitted to ApJ

Halpern, J.P. & Ruderman, M., 1993, ApJ, 415, 286

Harnden, F.R., Grant, P.D., Seward, F.D., & Kahn, S.M., 1985, ApJ, 299, 828

Koribalski, B., Johnston, S., Weisberg, J.M. & Wilson, W. 1995, ApJ, 441

Johnston, S., Manchester, R.N., Lyne, A.G., Kaspi, V.M. & D'Amico, N., 1994, to appear in MNRAS

McAdam, W.B., Osborne, J.L & Parkinson, M.L., 1993, Nat, 361,516

Michel, F.C., 1991, Theory of Neutron Star Magnetospheres, University of Chicago Press, ISBN 0-226-52331-4

Müther, H.,Prakash, M. & Ainsworth, T.L., 1987, Phys. Lett. B119, 469

Nomoto, K. & Tsuruta, S., 1987, ApJ, 312, 711

Page, D., 1994, ApJ (7/94)

Predehl, P. & Schmitt, J., 1994, accepted for publication in A&A

Ögelman, H., Finley, J.P. & Zimmerman, H.U., 1993, Nat, 361, 136

Ögelman, H.B., 1994, in *The Lives of Neutron Stars*, eds A. Alpar, U. Kiliźoğlu & J. van Paradijs, Kluwer Academic Publishers, in print

Safi-Harb, S. & Ögelman, H., 1994, in *The Lives of Neutron Stars*, eds A. Alpar, U. Kiliźoğlu & J. van Paradijs, Kluwer Academic Publishers, in press

Taylor, J.H., Manchester, R.N. & Lyne, A.G., 1993, ApJS, 88, 529

Taylor, J.H. & Cordes, J.M., 1993, ApJ, 411, 674

Thompson, D.J., et al., 1992, Nat, 359, 615

Thorsson, Prakash, M. & Lattimer, J.M., 1994, preprint

Ulmer, M.P., 1994, ApJS, 90, 789

Umeda, H., Shibazaki, N., Nomoto, K. & Tsuruta, S., 1993, ApJ, 408, 286

Van Riper, K.A., 1991, ApJS, 74, 449

Wiringa, R.B., Fiks, V. & Fabrocini, A., 1988, Phys. Rev. C38, 1010

Yancopoulos, S., Hamilton, T.T. & Helfand, D.J., 1994, ApJ 429, 832

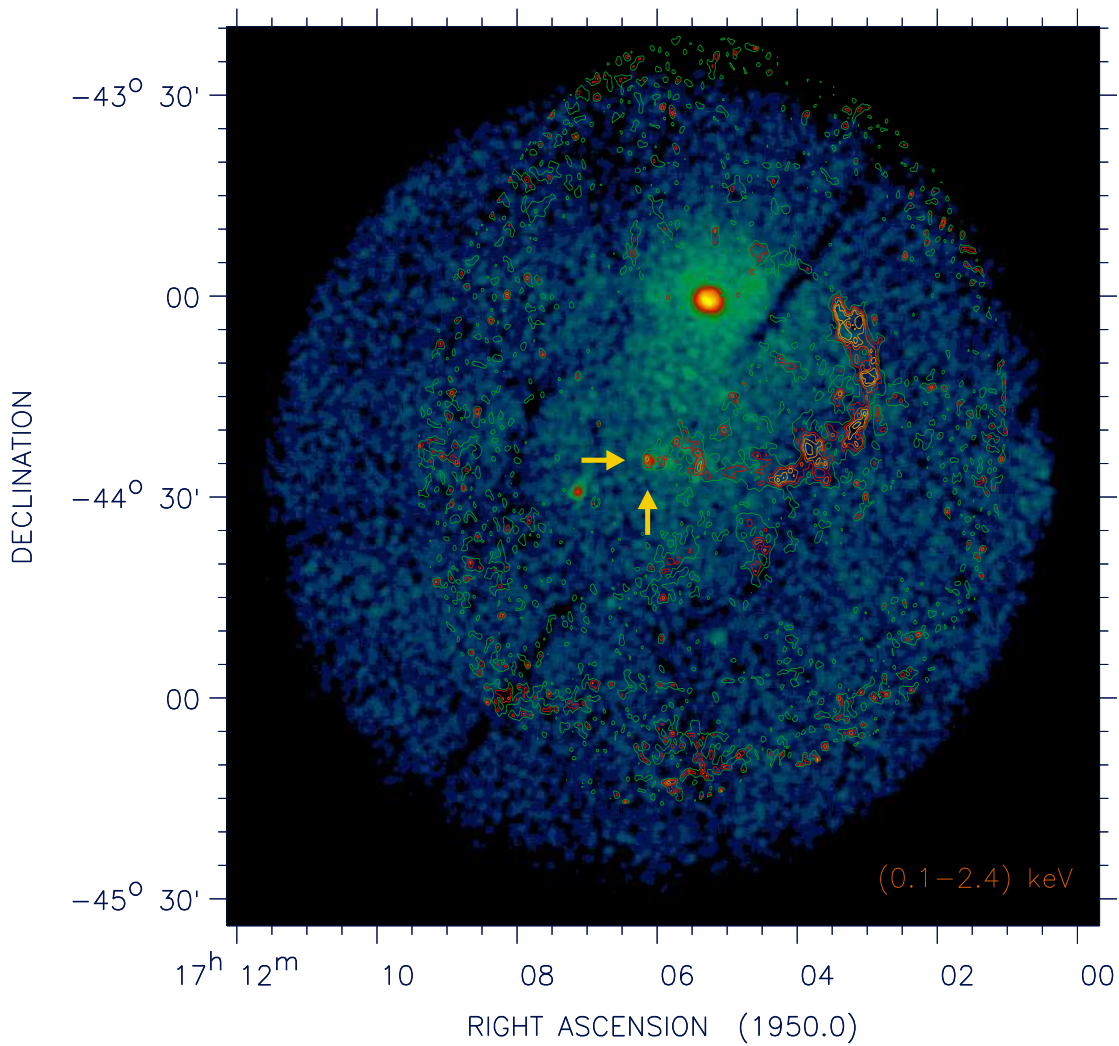

**Fig. 1.** Superposition of the 0.1-2.4 keV X-ray intensity map with the 843 MHz flux density map of the SNR G343-2.3 region (contour lines; radio point sources have been removed). The position of the pulsar at the edge of the elliptical half-shell is marked by arrows. Contour lines are at (2,4,5.5,7,9,28,32,200) mJy per beam of $43"\times 63"$. The bright X-ray source at the north is the low mass X-ray binary system 4U 1705-44. The contour lines cover the small X-ray source RX J1709.4-4429 close to the pulsar position. RX J1710.7-4433 is visible in south-east direction from the pulsar position.

to the detection of extended diffuse emission from the remnant is the presence of the strong X-ray source 4U 1705-44, which dominates a wide area of the PSPC field and may bury diffuse emission from the remnant.

To quantify the X-ray emission component from G343.1-2.3, we have computed a flux upper limit for that part of the supernova shell which is visible at 843 MHz with the Molonglo synthesis telescope (c.f. Fig. 1). The $2\sigma$ count rate upper limit determined for the radio shell is 0.018 cts/s. Assuming a power-law spectrum $dN/dE \propto E^{-\alpha}$ with photon-index $\alpha = 2$ and column density $N_H = 5 \times 10^{21}$ cm$^{-2}$, the count rate corresponds to an unabsorbed flux of $f_x \leq 9 \times 10^{-13}$ erg/s/cm$^2$ within $0.1-2.4$ keV.

## 4. Timing analysis

To test the X-ray flux for a modulation at the pulsar's radio frequency, a photon arrival time analysis was applied to the photons from within a radius of 60 arcsec from the source position for the February data and 50 arcsec for the September data, respectively. These radii were chosen with respect to the point spread functions and include more than 99.9% of the source photons. About 10% of the selected photons are background photons. Because of the large gap between the two ROSAT observations and the occurrence of a glitch between 3rd May and 9th July 1992 (Kaspi, private communication), the February and September data were analysed separately. Despite the